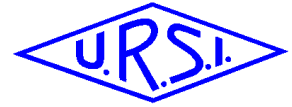

# Integration of Reconfigurable Intelligent Surfaces in Dynamical Energy Analysis

Sergio Terranova[(1)], Martin Richter[(1)], Neekar Mohammed[(1) (2)], Gabriele Gradoni[(1)], and Gregor Tanner [(1)]
(1) School of Mathematical Sciences, University of Nottingham, Nottingham, United Kingdom
(2) Department of Mathematics, College of Science, University of Sulaimani, Sulaymaniyah, Kurdistan Region, Iraq
(e-mail: sergio.terranova@nottingham.ac.uk, martin.richter@nottingham.ac.uk, neekar.mohammed2@nottingham.ac.uk
gabriele.gradoni@nottingham.ac.uk, gregor.tanner@nottingham.ac.uk )

## Abstract

Reconfigurable intelligent surfaces have been recently investigated for their potentials to offer significant performance improvements in the next generation wireless telecommunication systems (5G and beyond / 6G). Intelligent surfaces are programmed to control the electromagnetic propagation and obtain the desired wavefront by tuning the local reflection phase of unit elements. Predicting the electromagnetic propagation in the RIS-assisted wireless channel accurately is a significant challenge for researchers and becomes crucial for Telecom operators to properly allocate the radio resources. We propose the use of an Eulerian ray-tracing method, the Dynamical Energy Analysis (DEA), as a coverage planning tool capable of account for the EM interaction between reconfigurable intelligent surfaces and the surrounding environment. The main characteristics that make DEA suitable for this purpose are discussed and some preliminary results of the reflective surface integration within the DEA code will be presented.

## 1. Introduction

After the deployment of the fifth-generation (5G) wireless network, a large part of the scientific community is investing significant resources on the next generation communication system (6G), exploring promising technologies able to pave the way towards new human-centered smart-connected environments [1]. Often, one of the main disadvantages in conventional wireless communications is the uncertainty of the wireless propagation environment and its uncontrollability. In this context, electromagnetic metamaterials and reconfigurable intelligent surfaces (RISs) will serve as a new enabling technology to engineer the radio signal propagation in wireless networks. By properly tuning the local reflecting properties of the RIS unit cell, the electromagnetic wavefront can be engineered. Many low-cost passive reflecting elements are needed so that the RIS is able to dynamically control wireless channels and thus enhance the communication performance. Compared to other technologies, e.g., phased arrays and relays, RISs require more resonant elements. However, components such as PIN diodes and varactors are low cost, and they do not require radiofrequency power amplifiers. A self-consistent wave modelling approach accounting for the electromagnetic interaction between RISs and reflecting environments has not been explored in the literature. Indeed, in order to optimize the deployment of RISs in complex propagation environments, it is essential to correctly evaluate the RIS scattered field within a ray tracing coverage planning tool to incorporate the environmental complexity and the local behavior of RISs [2]. In this paper, we propose to adopt the Dynamical Energy Analysis (DEA) [3], an Eulerian ray-tracing method exploiting the so-called phase-space representation, in order to integrate the deployment of the RISs in complex wireless propagation environments. The rest of the paper is organized as follows: in Section 2 the DEA approach is briefly discussed; in Section 3 some preliminary numerical results are presented for a RIS-assisted environment, by adopting the DEA approach. Finally, some conclusions are drawn in the last section.

## 2. Dynamical Energy Analysis (DEA)

In order to predict the RIS-aided channel propagation in a large environment typical for an urban wireless scenario, a reasonably fast and accurate EM simulation method is needed. Two of the main asymptotic methods widely used in this context are Ray Tracing techniques (RT) [4], [5] and the Power Balance Method (PWB) [6]. When RT techniques are used to describe scenarios where one has to take into account multiple reflections, such as bounded domains i.e. indoor environments, the number of rays needed grows exponentially. Likewise, it can often be difficult to meet the assumptions underlying the PWB method, such as ergodicity, as well as the challenge of dividing the system into subsystems. DEA is a numerical energy flow method introduced in [3], which is able to compute densities of rays and transport them on numerical meshes. This method falls in between ray tracing and power balance methods, keeping the details of the environment typical of RT, while relaxing the PWB assumptions. DEA can be considered as a reformulation of RT in a joint space of position and direction of travel, referred to as phase-space, by using the Wigner function transformation. Because of this approach, by tracing the phase-space densities in this new computational domain, it captures bundle of rays instead of individual rays. The first step in obtaining a phase-space representation of wave-

fields is to define the Wigner distribution function (WDF) of a scalar field $\psi(x)$ as:

$$W(x,p) = \int \psi\left(x + \frac{u}{2}\right) \psi^*\left(x - \frac{u}{2}\right) e^{-ikpu} du$$

where x describes the position and p the local momentum. Similarly, one can define the WDF for a scalar function given in momentum representation $\hat{\psi}(p)$ as:

$$W(x,p) = \left(\frac{k}{2\pi}\right)^d \int \hat{\psi}\left(p + \frac{q}{2}\right) \hat{\psi}^*\left(p - \frac{q}{2}\right) e^{-ikqx} dq$$

where d is the dimension of the problem space. For example, if d = 1, $\hat{\psi}(p)$ is given by:

$$\hat{\psi}(p) = \int \psi(x) e^{-ikpx} dx$$

Now, consider for example the EM radiation from a complex EM source like a large Printed Circuit Board (PCB), where thousands of electrical devices are embedded in an enclosure [7]. A deterministic approach aiming to properly predict the radiation may be a very challenging task, hence we need to characterize such sources with statistical methods. Inherently, one can follow an approach based on measuring and propagating such statistically correlated wave field described by the field-field autocorrelations functions (ACFs), exploiting the WDF techniques. One then defines the two-point field-field correlation function as an ensemble average over different wave fields:

$$\Gamma(x_1, x_2) = <\psi(x_1)\psi^*(x_2)>$$

To represent such wave fields in phase-space, one has:

$$W(x,p) = \int \Gamma\left(x + \frac{u}{2}, x - \frac{u}{2}\right) e^{-ikpu} du$$

where we use the following coordinates:

$$x = \frac{x_1 + x_2}{2}, \quad u = x_1 - x_2$$
$$x_1 = x + \frac{u}{2}, \quad x_2 = x - \frac{u}{2}$$

As an example of a phase-space representation of a wave-field exploiting WDF, consider the case of propagation in a 2D domain defined by the x-z plane. We also assume that a scalar component of the electric and magnetic field at z = 0 is given by a double Gaussian pulse:

$$\psi(x) = \frac{1}{\sqrt{2\pi\sigma_x^2}} \left( exp\left(-\frac{(x-\mu_1)^2}{2\sigma_x^2}\right) + exp\left(-\frac{(x-\mu_2)^2}{2\sigma_x^2}\right)\right)$$

Where $\sigma_x$ defines to the width of the pulses and $\mu_1$ and $\mu_2$ the centers. A phase-space representation of such a field is reported in Fig. 1. In particular, (a) shows the correlation function and (b) the phase-space representation of such a wave-field.

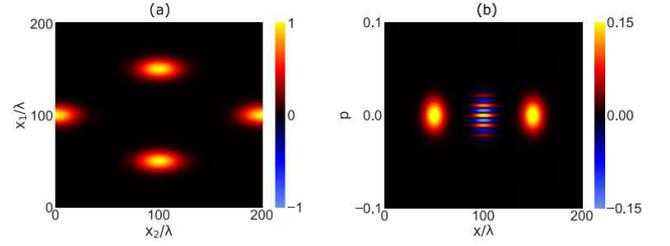

Fig. 1. (a) correlation function and (b) phase-space representation of the double Gaussian pulse wave-field.

This approach can be adopted also to represent the electromagnetic field scattered by a RIS, for example adopting a physical optics model and calculating the WDF of the scattered field upon a random plane wave incidence. While this approach is being pursued in a separate effort to derive the phase space scattering coefficients of a RIS, here we have focused on the numerical integration using DEA of reflective surfaces generating anomalous reflections.

## 3. Simulated Scenario

In this section, a simple application of the DEA ray-tracing method in a RIS-aided scenario is shown. The approach that we take is formulated in terms of ray densities in a dynamical phase-space, which represents rays in a joint space of average position and direction of travel (direction cosine of the wave vector). The method has been conceived to operate on meshes, so that also real-life environments can be studied. From a CAD model, a mesh-based representation of the environment under investigation can be generated to be compliant with the DEA solver. The meshes that can be used in DEA are those generated, for example, for finite element methods [8]. As shown in Fig. 2, a RIS is included in a defined polygonal meshed environment where the localized source, or the incident EM beam, is modeled as:

$$\rho_0(s,p) \sim cos^n\left(\frac{\pi}{2}(s - s_0)\right) cos^m\left(\frac{\pi}{2}(p - p_0)\right);$$

p represents the momentum, in the phase-space framework, and $s \in (0, L)$ is the coordinate along the boundary where L is the total length of the billiard boundary. In particular the RIS is modeled as an anomalous reflector and has been implemented into DEA code by the so-called Generalized Snell Law of Reflection and Diffraction as discussed in [9], where one has an additive term in addition to Snell's Law that depends on the gradient of the phase along the RIS. As is shown in Fig. 2, the RIS is located along the oblique edge of the polygon and showcase the effect of the anomalous reflection where we plot only one bounce of the beam at this RIS. However, in principle, also the other segments can take boundary conditions such as totally reflecting or absorbing, and carrying out an iterative computation, an equilibrium distribution can be achieved within this polygonal environment. Finally, to showcase this anomalous reflection behavior, in Fig. 3 a progressive phase-shift applied to the reflected beam is shown. It is obtained assuming a linear phase gradient imposed by the RIS in accordance with the Generalized Snell Law [9] .

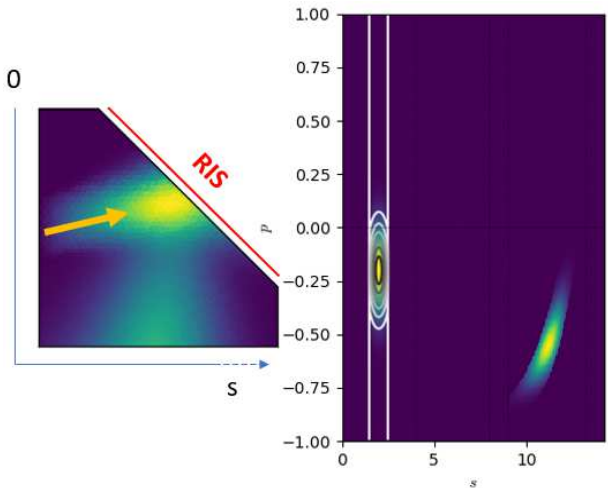

Fig. 2. On the left the spatial density in position space is shown, with s-coordinate moving along the perimeter of the polygon. On the right side the Boundary-map's phase space in (s, p) is represented.

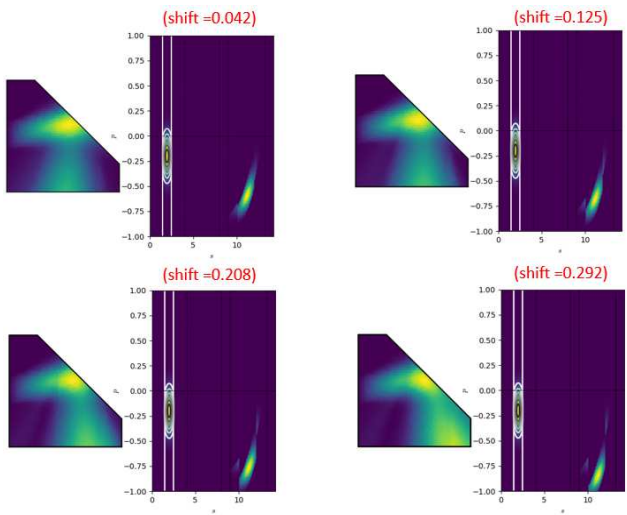

Fig. 3. Representation of the beam in the phase-space (s,p) by varying linearly the phase gradient of the RIS, for different values of the shift.

## 4. Conclusion

In this paper we have discussed the Dynamical Energy Analysis, a ray-tracing method based on high frequency asymptotic approximations able to propagate densities of rays in a meshed environment. In addition, the effect of the RIS, modelled as an anomalous reflector using the generalized Snell Law, has been showcased in some preliminary numerical results in a 2D environment by adopting the DEA approach. The results are very promising and pave the way for further developments. In the view of the authors, the next important development will be a phase-space representation of the field radiated by the RISs obtained using the Wigner Function technique and its full integration into DEA in a realistic 3D scenario.

## 5. Acknowledgements

This work has been supported by the European Commission by the H2020 RISE-6G Project under Grant 101017011.